\documentstyle[12pt]{article}
\def\be{\begin{equation}}
\def\ee{\end{equation}}

\def\bb{\begin{thebibliography}{99}\small\addtolength{\itemsep}{-6pt}}
\def\eb{\end{thebibliography}}

\makeatletter
\def\setb@se#1{\baselineskip=#1 \normalbaselineskip=#1}
\setb@se{12pt}
\long\def\title#1{\vspace*{11.5pc}{\pretolerance=10000\raggedright
  \setb@se{12pt}\bf #1\par}\nobreak\ignorespaces}
\long\def\author#1{\vspace{4pc}\begin{list}{\hfill}%
{\topsep=0pt\parskip=0pt\parsep=0pt\partopsep=0pt\listparindent=0pt%
\itemsep=0pt\rightmargin=0pt\labelsep=0pt\labelwidth=5pc\leftmargin=5pc}%
\item\normalsize{#1}\end{list}\vspace{14pt}}
\long\def\affil#1{\begin{list}{\hfill}%
{\topsep=0pt\parskip=0pt\parsep=0pt\partopsep=0pt\listparindent=0pt%
\itemsep=0pt\rightmargin=0pt\labelsep=0pt\labelwidth=5pc\leftmargin=5pc}%
  \item\normalsize{\rm #1}\end{list}\vspace{7pt}}
\long\def\beginabstract{\vspace{21pt plus 7pt minus 7pt}\begin{list}{\hfill}%
{\topsep=0pt\parskip=0pt\parsep=0pt\partopsep=0pt\listparindent=0pt%
\itemsep=0pt\rightmargin=0pt\labelsep=0pt\labelwidth=5pc\leftmargin=5pc}%
\item\normalsize{\bf Abstract. }}
\long\def\endabstract{\end{list}\vspace{28pt plus14pt minus 14pt}%
\normalsize\noindent}
\def\@sect#1#2#3#4#5#6[#7]#8{\ifnum #2>\c@secnumdepth
  \def\@svsec{}\else
  \refstepcounter{#1}\edef\@svsec{\csname the#1\endcsname.\hskip 0.5em}\fi
  \@tempskipa #5\relax
   \ifdim \@tempskipa>\z@
     \begingroup #6\relax
       \@hangfrom{\hskip #3\relax\@svsec}{\interlinepenalty \@M #8\par}%
     \endgroup
    \csname #1mark\endcsname{#7}\addcontentsline
      {toc}{#1}{\ifnum #2>\c@secnumdepth \else
		   \protect\numberline{\csname the#1\endcsname}\fi
		 #7}\else
     \def\@svsechd{#6\hskip #3\@svsec #8\csname #1mark\endcsname
		   {#7}\addcontentsline
			{toc}{#1}{\ifnum #2>\c@secnumdepth \else
			  \protect\numberline{\csname the#1\endcsname}\fi
		    #7}}\fi
  \@xsect{#5}}

\def\section{\@startsection {section}{1}{\z@}{-28pt plus -14pt minus
-14pt}{2.3ex plus .2ex}{\normalsize\bf}}
\def\subsection{\@startsection{subsection}{2}{\z@}{-14pt plus -8pt
minus -4pt}{1.5ex plus .2ex}{\normalsize\bf}}
\def\subsubsection{%
    \@startsection{subsubsection}{3}{\z@}{-14pt plus
    -8pt minus -4pt}{-1.5ex plus -.2ex}{\normalsize\bf}}
\def\paragraph{\@startsection
     {paragraph}{4}{\z@}{3.25ex plus 1ex minus .2ex}{-1em}{\normalsize\bf}}
\def\subparagraph{\@startsection
     {subparagraph}{4}{\parindent}{3.25ex plus 1ex minus
     .2ex}{-1em}{\normalsize\bf}}

\def\caption{\refstepcounter\@captype \@dblarg{\@caption\@captype}}
\long\def\@caption#1[#2]#3{\par\addcontentsline{\csname
 ext@#1\endcsname}{#1}{\protect\numberline{\csname
 the#1\endcsname}{\ignorespaces #2}}\begingroup
   \@parboxrestore
   \hspace*{28pt}
   \parbox{394pt}{\@makecaption{{\bf\csname
	    fnum@#1\endcsname}}{\ignorespaces #3}}\par
\vspace{7pt}\endgroup}

\long\def\@makecaption#1#2{
   \vskip 14pt
   \setbox\@tempboxa\hbox{#1{\bf.} #2}
   \ifdim \wd\@tempboxa >\hsize   
       #1{\bf.} #2\par            
     \else                        
       \hbox to\hsize{\box\@tempboxa\hfil}
   \fi}

\newcommand{\boldarrayrulewidth}{1pt} 

\def\bhline{\noalign{\ifnum0=`}\fi\hrule \@height
		    \boldarrayrulewidth \futurelet
		    \@tempa\@xhline}

\def\@xhline{\ifx\@tempa\hline\vskip \doublerulesep\fi
      \ifnum0=`{\fi}}

\def\thebibliography#1{\section*{REFERENCES\@mkboth
  {REFERENCES}{REFERENCES}}\list
  {\hfil[\arabic{enumi}]}{\itemsep=0pt\labelsep=7pt\itemindent=-14pt
    \settowidth\labelwidth{[#1]}
    \leftmargin\labelwidth
    \advance\leftmargin\labelsep
    \advance\leftmargin -\itemindent
    \usecounter{enumi}}\setb@se{12pt}\small
    \def\newblock{\hskip .11em plus .33em minus .07em}
    \sloppy\clubpenalty4000\widowpenalty4000
    \sfcode`\.=1000\relax}
\def\references{\section*{REFERENCES\@mkboth
{REFERENCES}{REFERENCES}}\list{}{\itemsep=0pt\labelsep=0pt\itemindent=-28pt
\labelwidth=0pt\leftmargin=28pt}\setb@se{12pt}\small
\def\newblock{\hskip .11em plus .33em minus .07em}
\sloppy\clubpenalty4000\widowpenalty4000
\sfcode`\.=1000\relax}
\newcommand{\etal}{{\em et al\/}\ }

\makeatother

\newcommand{\beq}{\begin{equation}}
\newcommand{\eeq}{\end{equation}}
\newcommand{\bq}{\begin{quotation}}
\newcommand{\eq}{\end{quotation}}
\newcommand{\bc}{\begin{center}}
\newcommand{\ec}{\end{center}}

\catcode `\@=11
\def\journal#1&#2&#3(#4){\begingroup \let\journal=\dummyj@urnal
    \unskip~\sl #1\unskip~\bf\ignorespaces #2\rm
    \unskip,~\ignorespaces #3
    (\afterassignment\j@ur \count255=#4)\endgroup\ignorespaces }
\def\j@ur{\ifnum\count255<100 \advance\count255 by 1900 \fi
	  \number\count255 }
\def\dummyj@urnal{%
    \toks@={Reference foul up: nested \journal macros}%
    \errhelp={You forgot & or ( ) after the last \journal}%
    \errmessage{\the\toks@ }}
\catcode `\@=12

\def\ltwid{\raise.3ex\hbox{$<$\kern-.75em\lower1ex\hbox{$\sim$}}}
\def\gl{\raise.5ex\hbox{$>$}\kern-.8em\lower.5ex\hbox{$<$}}
\def\gtwid{\raise.3ex\hbox{$>$\kern-.75em\lower1ex\hbox{$\sim$}}}

\def\3he{$^3\hbox{He}$}
\def\4he{$^4\hbox{He}$}

\def\eg{{\it e.g.}}
\def\ie{{\it i.e.}}
\def\a0{\text{ \AA}}

\def\etal{{\it et al.}}

\def\pd{\partial}

\def\bold#1{{\bf #1}}
\def\VEC#1{\bold{#1}}
\def\a0{\hbox{ \AA}}

\begin{document}


\title{DYNAMICS OF COSMOLOGICAL PHASE TRANSITIONS: WHAT CAN WE LEARN
FROM CONDENSED MATTER PHYSICS?\footnote{Invited lectures, to appear in
the proceedings of the NATO ARW on ``Formation and Interactions of
Topological Defects'', A.C. Davis and R.H. Brandenberger (eds.). Plenum
Press., N.Y., 1995.}}

\author{Nigel Goldenfeld\footnote{e-mail: {\tt nigel@uiuc.edu}}}

\affil{Department of Physics, University of Illinois at
Urbana-Champaign, \\ 1110 West Green St., Urbana, Il. 61801-3080,
U.S.A.}

\beginabstract
A brief outline is given of the description of phase transition kinetics
in condensed matter systems with a continuous symmetry, emphasising the
roles of dissipation, coarse-graining and scaling.  The possible
relevance of these ideas to the early universe is explored in the
contexts of the GUT string transition and the electroweak transition.
\endabstract

\section{INTRODUCTION}

\noindent
How fast do phase transitions occur?  Remarkably, it is found that
although it is often straightforward to estimate a characteristic
relaxation time for the microscopic degrees of freedom, the actual
characteristic time for completion of the phase transition may be many
orders of magnitude greater.  For example, laboratory experiments
indicate that following a temperature quench, the transition to the
superconducting state of a normal metal in a magnetic field may take
many minutes.  The primary reason for the slowness of the transition is
the formation, interaction and subsequent dynamics of topological
defects.  In systems with a discrete symmetry and a scalar order
parameter, such as binary alloys, the topological defects are domain
walls, whilst in systems with a continuous symmetry and a vector or
tensor order parameter, such as certain liquid crystals, the defects may
be strings and monopoles.  The motion and mutual annihilation of the
defects is usually the rate-determining step for the transition, and is
affected by such factors as the presence of dissipation or disorder, the
range of the interactions between defects, and even their homotopy
classification.

Such phenomena are ubiquitous in condensed matter, yet only recently
have detailed studies been made of phase transition kinetics in
condensed matter systems with non-scalar order parameters.  Although
interesting in their own right, these examples may be regarded as
caricatures of the phase transitions that are thought to have occurred
in the early universe as it cooled.  The principal analogous feature is
the spontaneously broken global continuous symmetry and, in
superconductors at least, the existence of a local gauge symmetry.

The purpose of this article is to outline briefly the way in which the
kinetics may be investigated theoretically, with an eye towards using
the techniques of condensed matter physics in a cosmological context,
where appropriate.  In particular, the work of the group at Illinois on
phase ordering in systems with continuous symmetries is relevant.  This
primarily numerical work includes studies of the non-conserved dynamics
of the XY model in two dimensions\cite{mg}, three dimensions\cite{mg3},
the conserved XY model in one and two dimensions\cite{mgcon}, the
dynamics of the superconducting transition\cite{lmg}, the dynamics of
the Ising gauge theory, where there is no local order
parameter\cite{liu} and ordering in uni- and bi-axial liquid
crystals\cite{zgg}.  Other relevant numerical studies are those of
Toyoki\cite{toynum} and Bray, who has recently given a complete review
of the topic of phase ordering in systems with continuous
symmetries\cite{brayrev}.

Since all of the literature and an extensive review are easily available,
it does not seem worthwhile to provide duplication of the
results here.  Instead, the focus will be on principles and concepts,
together with some remarks comparing the procedures used in condensed
matter physics with those used in cosmological applications.  These are
particularly pertinent in the case of the electroweak transition, where
the phase transition kinetics may not occur via nucleation and growth,
as has sometimes been assumed.

\subsection{Scaling}

Let us consider qualitatively the sequence of different time regimes
exhibited by a binary alloy undergoing spinodal decomposition.  First,
the small amplitude long-wavelength fluctuations of the order parameter,
present from the initial conditions, become amplified exponentially:
this behaviour is predicted by the linearised equations of motion, and
is rarely observed.  The nonlinearities in the equation of motion
quickly stabilise the order parameter at its two equilibrium values
almost everywhere.  Thus a series of domains has been formed, separated
by domain walls.  This interlocking pattern of domains subsequently
coarsens, driven by the excess energy from the curvature of the domain
walls.  At very long times, the system attains equilibrium, which
energetically should simply be a single domain wall dividing the system
into two coexisting equilibrium phases.  The intermediate time regime,
where the domains simply coarsen, exhibits  {\it dynamic
scaling\/}\cite{reviews}.

Dynamic scaling simply means that at large enough times $t$, the
emerging pattern contains only one time-dependent length scale $L(t)$.
Thus, the equal time two-point correlation function of the order
parameter $\psi$ defined by
\begin{equation}\label{scalinglaw}
C({\bf r}, t) \equiv \left<\psi({\bf r},t)\psi({\bf 0},t)\right>
\end{equation}
is actually only a function of a reduced variable:
\be\label{reduced}
C({\bf r}, t)=F(r/L(t))
\ee
where $F$ is known as a scaling function.  In practice, for finite
times, there is a weak explicit time dependence in $F$, which proceeds to
increasingly short distances, whilst $L(t)$
converges to its \lq ideal'
power law form, as discussed below.

Of course, there are other length scales present: the bulk correlation
length for longitudinal order parameter fluctuations $\xi (T)$, which is
a function of the temperature $T$ after the quench, and the microscopic
length scale $a$, on the order of molecular dimensions.  But for
sufficiently long times, the following inequalities hold: $L(t) \gg \xi
(T) > a$.  Since thermal fluctuations are operative up to the scale of
$\xi $, the unimportance of $\xi$ relative to $L$ is sometimes stated in
a suggestive way by saying that phase ordering \lq\lq is controlled by a
zero temperature fixed point".  (Despite the intuition that some sort of
renormalisation group (RG) approach should form the basis of a general
theory for the approach to equilibrium, a predictive RG theory remains
elusive.)  The scaling regime may be called an {\it intermediate
asymptotic regime}.  For example, in a binary alloy, at short times, the
initial fluctuations in the system are being amplified by the unstable
growth process, and domain walls are formed, whereas for long enough
times in a finite system, the system reaches thermal equilibrium, where
only one domain wall traverses the system.  In both of these regimes,
thermal fluctuations are important.  For some set of intermediate times,
whose duration is an increasing function of the system size, the scaling
regime is observed, and thermal fluctuations are not important in the
sense described above.  Empirically, it is found that the growth of the
characteristic scale $L$ follows a power law form:
\be \label{powerlawform}
L(t)= A \xi (T)\,
(t/\tau(T))^{\phi},
\ee
where $\tau (T)$ is the temperature dependent order parameter relaxation
time, $A$ is an amplitude assumed to be of order unity (but not yet
measured as far as I know), believed to be universal and $\phi$ is now
believed to have the value $1/3$ for alloys or other conserved systems
with a discrete symmetry in dimension $d \ge 2$.  For systems with a
non-conserved parameter in these dimensions, $\phi$ has the value
$1/2$.  Eq. \ref{powerlawform} is valid near the transition temperature
$T_c$, and a more general form valid for all $T< T_c$ has been proposed
by Bray\cite{bray90}.

Heuristic arguments, given elsewhere in this volume (see, for example,
the lectures by Bray), connect these power laws with the motion of
domain walls.  So one is led to ask how these scaling results are
affected when the system exhibits a continuous symmetry, rather than a
discrete one, so that the domain walls no longer exist. In fact, other
topological defects are present in such systems, and can give rise to
different growth laws.  These considerations formed the motivation for
the work done at Illinois.

\subsection{Computer simulation}

In order to observe quantifiably the scaling regime in a computer
simulation, two conditions must be met: the system size needs to be as
large as possible, so that the regime lasts as long as possible, and
then the longest possible times must be attained.  In addition to these
requirements, a number of other results are known which are diagnostics
of the scaling regime: that is, they are only satisfied in the scaling
regime.  These are the Tomita sum rule\cite{tomita}, Yeung's law for the
$k\rightarrow 0$ behaviour of the X-ray scattering form factor at small
wavenumbers $k$\cite{yeung}, and Porod's law for the form factor's short
distance behaviour\cite{porod}.  In the absence of an obvious small
parameter, systematic analytical work has rarely been possible --- a
notable exception is the study of the $O(N)$ model for
$N\rightarrow\infty$, where there are no topological
defects\cite{coniglio} --- and much of our knowledge has come from
computer simulations.  Thus, in order to estimate accurately (\eg) the
exponent $\phi$, computationally efficient techniques have been designed
to probe as far into the scaling regime as possible: these are discussed
in section 3.3.  In fact, rather little of the numerical work to date
actually satisfies the known criteria for the asymptotic regime; the
most complete work to date on the alloy phase separation problem in
three dimensions is that of Shinozaki and Oono\cite{shin}.

\section{FORMULATION}

\subsection{Level of description}

The main difficulty in constructing a theory for phase ordering is the
disparity between the different length scales present ($\xi$, $L(t)$ and
$a$), and the complexity of the actual microscopic equations of motion.
However, we are mainly interested in phenomena on the scale of $L$, so a
coarse-grained description is adequate.  This sort of approach is common
in condensed matter physics: for example, the BCS theory of
superconductivity is a well-tested microscopic theory but is virtually
useless in situations with spatial variation, such as near boundaries or
for time dependent phenomena.  Instead, the Ginzburg-Landau theory, a
phenomenological theory for coarse-grained order parameter, slowly
varying on the scale of $\xi$, is used.

\subsection{Coarse-grained order parameter}

Let us be more precise, taking as our example, the case of a
superconductor.  The order parameter $\Psi$ is zero for $T>T_c$ and
nonzero for $T<T_c$.  It can be defined in terms of an anomalous Green
function by
\be\label{opdef}
\Psi (\VEC x) \propto \left<\hat \psi_{\downarrow} (\VEC x)
\hat \psi_{\uparrow} (\VEC x)\right>,
\ee
where $\hat\psi_{\downarrow} (\VEC x)$ is a down-spin electron field
operator and the angle brackets denote an equilibrium thermal
expectation value\cite{fetter}.  The normalisation of $\Psi$ may be
chosen with a convention that does not concern us here.  In the Meissner
and normal phases of a superconductor, the system is translationally
invariant and $\Psi$ is spatially uniform.  Near equilibrium, $\Psi$ may
vary in space, but we will only consider the long wavelength
variations:  in spinodal decomposition, the instabilities occur at
wavelengths long compared with $a$.  Conceptually, the coarse-grained
order parameter is defined by the long wavelength Fourier components
$\tilde \Psi_{\VEC k}$:
\be\label{coarsedef}
\Psi_\Lambda (\VEC x) \equiv \sum_{|\VEC k| <
\Lambda}e^{i\VEC k\cdot\VEC x} \tilde \Psi_{\VEC k}
\ee
where the coarse-graining scale $\Lambda$ satisfies $a \ll \Lambda^{-1} < \xi$.
We want to coarse-grain as much as possible, but if we coarse-grain
beyond the correlation length then we will have two-phase coexistence
within one coarse-graining volume.  This over-coarse-grained description
would then not be able to tell us about the dynamics of phase
separation.  The coarse-grained free energy governs the effective
dynamics of these long wavelength modes, and is obtained by integrating
out the short wavelength modes up to the scale $\Lambda$:
\be\label{coarsefree}
e^{-F_\Lambda\{\Psi_\Lambda (\VEC x)\}/k_BT}
\equiv \int \prod_{|\VEC k| > \Lambda} d\Psi_{\VEC k} e^{-H/k_BT}
\delta \bigl(\Psi_\Lambda (\VEC x) - \sum_{|\VEC k| <
\Lambda}e^{i\VEC k\cdot\VEC x} \tilde \Psi_{\VEC k}
\bigr),
\ee
where $H$ is the Hamiltonian and $k_B$ is Boltzmann's constant.
Thus, for a given coarse-grained order parameter profile in space, a
value for the coarse-grained free energy can be calculated.  In thermal
equilibrium, the probability that the coarse-grained order parameter has
a given profile $\Psi_\Lambda (\VEC x)$ is proportional to $\exp \left(
-F_\Lambda\{\Psi_\Lambda (\VEC x)\}/k_BT\right)$.  The explicit
calculation of $F_\Lambda$ is technically complicated, and unnecessary
to perform if one is interested in universal equilibrium quantities.  If
one is interested in computing the dynamics, certain aspects may not be
universal, and will depend upon $F_\Lambda$.  A recent example where
$F_\Lambda$ has been explicitly calculated using the RG is the work of
Alford and March-Russell\cite{alford}.

It is important to notice the distinction between the calculation of
$F_\Lambda$ and the finite temperature effective action $S$ of field
theory.  The latter involves no notion of coarse-graining (\ie\ $\Lambda
=\infty$), and is rigorously convex.  On the other hand, there is no
requirement of convexity for the coarse-grained free energy, and indeed,
it has the familiar double-well or wine-bottle form for $T<T_c$.  The
non-convexity of the one-loop effective action is an artifact of the
loop expansion.

\subsection{Dynamics}

How should we describe the dynamics of the phase transition?  Ideally
one would solve for the time-dependent density matrix, using the exact
microscopic Hamiltonian.  Observables such as the coarse-grained order
parameter, or its correlation functions, could then be computed.  This
would involve averaging over the initial conditions with the appropriate
Boltzmann weight, and modelling the quench itself.  In other words, one
would compute the exact dynamics, then coarse-grain.  An alternative
procedure would be to find the effective equations of motion for the
coarse-grained order parameter by directly coarse-graining the equations
of motion for the density matrix.  A plausible substitute is to write
down the phenomenological equation of motion for $\Psi_\Lambda$,
assuming that the driving force is $F_\Lambda$.  Even this is rarely
done, because usually $F_\Lambda$ is not known.  Instead, a
phenomenological form of $F_\Lambda$ is used.

A phenomenological Langevin equation for $\Psi_\Lambda$ is obtained by
assuming that the force driving the system towards thermal equilibrium
is proportional to the deviation from equilibrium:
\be\label{phenmotion}
\tau_0\pd_t\Psi_\Lambda = - \frac{\delta F_\Lambda}{\delta
\Psi_\Lambda^*}
+ \eta
\ee
where we have assumed that $\Psi_\Lambda$ is a complex scalar field
(appropriate for a superfluid, for example), and included a thermal
noise term $\eta$ so that the system is guaranteed eventually to attain
the global minimum of $F_\Lambda$.   We shall refer to eq.
\ref{phenmotion} as the time-dependent Ginzburg-Landau equation (TDGL).
As mentioned above, the noise term is believed not to play a significant
role during the scaling regime.  The relaxation time $\tau_0$ is in
principle calculable from a microscopic theory (such as BCS theory in
the case of a superconductor).  The Langevin equation above does not
include any conservation laws, although these are easy to include when
required.  For a superconductor, an appropriately generalised local
gauge invariant time-dependent Ginzburg-Landau description can be
written down (see \eg, \cite{lmg}).

It is difficult to assess the regime of validity of eq.
\ref{phenmotion}.  Usually it is regarded as a minimal model of phase
ordering kinetics, in the sense that it correctly predicts all universal
phenomena.  However, it may not be accurate for other quantities, and in
this sense, it is sometimes said that eq.  \ref{phenmotion} is a
semiquantitative description.  An important point is that we have
assumed that the dynamics is purely relaxational.  Whilst this is
expected to be valid near a critical point, where the correlation length
is large, for quenches to low temperatures this is not necessarily the
case.

Let us use again our example of the coarse-grained dynamics of a
superconductor to see what can go wrong.  There, the microscopic BCS
theory may be formulated as a set of coupled equations for the Green's
functions of the theory, which in turn can be reexpressed as an integral
equation for the order parameter\cite{fetter}.  The TDGL is obtained by
expanding the integral equation in powers of wavenumber $k$ and
frequency $\omega$.  However, only at low temperatures, near the
critical temperature or when there is strong scattering from impurities
does this procedure yield a result independent of the {\it ratio\/}
$\omega/k$, leading to a {\it local\/} partial differential equation.
In all other situations, there is no possible local description of even
the long wavelength, low frequency order parameter dynamics.
Furthermore, at low temperatures, it turns out that the dynamics is not
overdamped but exhibits wave-like solutions\cite{wert}.

\section{NUMERICAL METHODS}

\subsection{Monte Carlo Simulation}

This is probably the most straightforward conceptually.  Here, the
Monte Carlo time is assumed to be proportional to real time, and the
dynamics that the system undergoes is assumed to be somehow similar to
that of the real system.  Of course, this may not be the case: Monte
Carlo simulation is only a stochastic process which samples the correct
equilibrium distribution.  However, a judiciously chosen Monte Carlo
dynamics may well be a reasonable caricature of the actual dynamics.
This method is usually very slow, and it has not proved to be reliable
in extracting the correct long time behaviour.  It is not the method of
choice nowadays.

\subsection{Molecular Dynamics for the Defects}

As discussed at length in this volume, topological defects are created
during the phase transition, and equilibrium may only occur after all
such defects have annihilated.  In many situations, the potential for
interactions between defects may be calculated, and the defects treated
as classical particles, subject to damping and the inter-defect
potential.  Accordingly the equation of motion for the particles may be
readily solved.  An example, in the case of superconductor dynamics, is
ref. \cite{enomoto}.  This approach is relatively fast in terms of
computer time, and large systems may be straightforwardly treated.  The
disadvantage, however, is that it is usually non-trivial to obtain the
correct equation of motion for the defects, either because the potential
may be hard to obtain, or because the damping may be difficult to
calculate.  An interesting example is the calculation of ordering in the
non-conserved two dimensional XY model in ref. \cite{toyoki}.  Here, the
ordering proceeds via the annihilation of $\pm$ vortices, whose
interaction potential $U$ is logarithmic in their separation $R$.  The
equation of motion for the separation $R$ of two vortices is assumed to
be
\be\label{sep}
\nu \frac{dR}{dt} = - \frac{dU}{dR}
\ee
where $\nu$ is a damping coefficient.  This calculation, as well as the
molecular dynamics simulation with many vortices, finds that the average
separation $L$ varies as $L\sim t^{1/2}$, whereas the correct result is
now believed to be $L\sim (t/\log t)^{1/2}$: the logarithmic factor
arises because the damping coefficient $\nu$ is actually logarithmically
dependent on $R$ \cite{pleiner}.  Note that the above estimate of
scaling exponents, based upon the overdamped equation of motion for a
pair of topological defects, works well in other situations, even
correctly predicting the crossover in a superconductor with penetration
depth $\lambda$ from $L\sim (t/\log t)^{1/2}$ ($L<\lambda$) to $L\sim
\log t$ ($L>\lambda$)\cite{lmg}.

\subsection{Order Parameter Evolution - PDEs and CDS}

Probably the most effective way to explore phase transition kinetics is
to solve directly the TDGL or equivalent partial differential equation
(PDE).  This involves discretising the PDE on a space-time lattice:
\be\label{discr}
\Psi ({\bf x},t)\rightarrow \Psi ({\bf n} \Delta x, i \Delta t)
\equiv \Psi^{i}_{\bf n}
\ee
and using a time-stepping algorithm such as the explicit Euler scheme
\be\label{euler}
\Psi_{\bf n}^{i+1} = \Psi^{i}_{\bf n} +
\Delta t \, \left[-\frac{\delta F}{\delta \Psi^*}\right]^i_{\bf n}.
\ee
Of course, eventually the continuum limit must be taken: $\Delta x$,
$\Delta t\rightarrow 0$, subject to possible stability criteria giving
an upper bound to $\Delta t/\Delta x^2$.  The disadvantage of this
method is primarily that of speed and memory: to explore asymptotically
large times, with a time step tending to zero requires many iterations
of eq. \ref{euler}.

The cell dynamic system (CDS) method\cite{oonopuri} exploits
universality in order to overcome this problem, and is now the most
widely used approach in studying phase ordering and other pattern
formation problems in condensed matter.  The basic idea is the
observation that it is rather wasteful to model the phenomenon in
question by a PDE, which one must then discretise to obtain a set of
coupled maps (such as eq. \ref{euler}) suitable for numerical
computation.  Instead, the phenomenon is modelled directly by a set of
coupled maps, defined on a coarse-grained space-time lattice,
with spatial cells of dimension of order $2\pi\Lambda^{-1}$.

To illustrate the basic idea, consider the case of an order-disorder
transition, where the order parameter is a simple non-conserved real
scalar field obeying the TDGL, and equilibrating with the coarse-grained
free energy
\be\label{odfe}
F \{\Psi({\bf x})\} = \int d^d{\bf x}\, \left[ \frac{1}{2}
(\nabla\Psi)^2 +
\frac{a}{2}\Psi^2 + \frac{b}{4}\Psi^4\right].
\ee
The parameter $a$ changes sign at $T_c$, is initially positive when
$T>T_c$, and at $t=0+$ is supposed to become instantaneously negative.
When substituted into the TDGL, a nonlinear PDE is obtained.  The CDS
approach need make reference neither to the TDGL nor to the
coarse-grained free energy functional of eq. \ref{odfe}.  Let us
consider the dynamics of the order parameter in one cell, throughout
which the order parameter value is essentially constant, but allowed to
vary in time.  The time dependence of $\Psi$ will have three fixed
points, two symmetrically placed about one at $0$, corresponding to the
three extrema at $\Psi=0$, $\pm\sqrt{a/b}$.  The former is an unstable
fixed point of the cell dynamics, whereas the other two are stable fixed
points.  We can model the cell dynamics by {\it any\/} map ${\cal
M}_{\Delta t}$ that has this fixed point structure.  Thus
\be\label{cdsi}
\Psi_{\bf n}^{i+1} = {\cal M}_{\Delta t} \{\Psi^{i}_{\bf n}\}.
\ee
It is often convenient to take the map
\be\label{mape}
{\cal M}_{\Delta t}\{\Psi\} = 1.3 \tanh (\Psi),
\ee
although even simpler piece-wise linear maps have also been used.  If
desired, one could explicitly calculate ${\cal M}_{\Delta t}$ from the
TDGL, although there would be no point for present purposes: the
phenomenological differential equation or phenomenological PDE have no
more privileged status than our phenomenological map.  The interaction
between cells should reflect the role of diffusion processes, and should
be as isotropic as possible.  Since diffusion is simply a local
averaging procedure, we couple the cells in the final form of our CDS by
writing
\be\label{mapfinal}
\Psi_{\bf n}^{i+1} = {\cal M}_{\Delta t} \{\Psi^{i}_{\bf n}\} +
D\left( \left<\Psi^{i}_{\bf n}\right> - \Psi^{i}_{\bf n}\right)
\ee
where $D$ is a phenomenological coefficent, akin to a diffusion
constant (but incorporating the spatial and time discretisation units)
and the averaging operation is in (\eg) two dimensions
\be\label{diffusion}
\left<\Psi^{i}_{\bf n}\right> \equiv \frac{1}{6}\sum_{n.n.}
\Psi^{i}_{\bf n}+ \frac{1}{12}\sum_{n.n.n.}\Psi^{i}_{\bf n}.
\ee
Here n.n. means nearest neighbours, n.n.n. means next nearest
neighbours.  This form of Laplacian is more isotropic than the
conventional discretisation\cite{shin,tom2}, as can be seen
by examining isocontours of its Fourier-transform.

Empirically, this CDS modelling and its extensions works very well in a
wide variety of situations, and gives results that in some cases have
been compared with direct integration of the TDGL.  This is not
surprising: the actual form of the map used is of little consequence for
large scale structure, and only influences the detailed form of the
order parameter profile near a domain wall or topological defect.  Of
course, since the free energy of eq. \ref{odfe} is only
phenomenological, its predictions for this variation are no more
reliable than those of the CDS map.

The key point about the CDS method, and one that is frequently
misunderstood, is that there is no continuum limit:  $\Delta x$ and
$\Delta t$ are not infinitesimals.  Thus, very rapid simulations are
possible on large systems.  The correct way to view the CDS map is that
one has integrated or coarse-grained some microscopic equations of
motion up to the space-time scale of interest, thus obtaining the map we
have guessed phenomenologically.  This is conceptually different from
the usual approach of numerical analysis, which is to sample a PDE on a
sequence of finer and finer meshes.  There is no proof that the CDS
algorithm is in the same universality class of the PDE (whatever that
may be), but the important point is that both are in the same
universality class as the physical phenomenon of interest.  This is the
job of the physicist: to identify and characterise such universality
classes.  Ref. \cite{rg} describes recent work on the application
of RG to extract universal features of PDEs.

In the literature, a popular variant on the CDS method is often
encountered, in which the conventional Euler discretisation of a PDE is
used, but with large $\Delta t$ and $\Delta x$.  This works well too,
although one must then be careful not to identify the potential used in
this discretisation of the PDE with the potential in the original
continuum limit PDE, for the reasons described in the preceding
paragraph.  The tanh map advocated above is simply a convenient form of
the potential, which happens to avoid a secondary numerical instability
by virtue of being injective.

\section{COSMOLOGICAL PHASE TRANSITIONS}

In this section, I will present some remarks and observations on several
issues of interest to cosmologists.  These arose during discussions at
the workshop, whilst presenting the procedures and insights obtained
from simulating string formation, evolution and the dynamics of local
gauge theories in condensed matter, and from inspection of computer
animations of phase transition kinetics with non-scalar order parameters.

\subsection{Violations of scaling}

Although scaling is observed generically in the presence of a
conservation law, the nonconserved order parameter dynamics only leads
to scaling solutions for symmetric or critical quenches.  Thus, if the
order parameter obeys the TDGL with the coarse-grained free energy of
eq. \ref{odfe}, then a symmetric or critical quench is one where the
spatial average $\left<\Psi({\bf x}, 0)\right>=0$ initially.  However,
it is not obvious that in a cosmological context, the ordering field
will always satisfy this condition.  For example, in a GUT transition at
which strings are formed, earlier symmetry breaking transitions may
create a bias in the Higgs field.  In the toy-model case of the three
dimensional XY model, simulations\cite{mg3} confirm the prediction from
mean field theory\cite{toymean} that scaling is violated, with the
initial string distribution rapidly breaking up into small loops.  The
total length of string $\ell(t)$ a time $t$ after the quench is given by
\be\label{stringlength}
\ell(t) \sim t^{-1}\exp(-b^2t^{3/2})
\ee
where the bias is given by $b^2 = |\left<\Psi({\bf x}, 0)\right>|^2$.
For times less than the crossover time $t_c \sim b^{-4/3}$, the dynamics
of the string network resembles that of the critical quench, but for
longer times, it becomes readily apparent that there are two length
scales present: the constant mean size of small loops, and the
increasing mean distance between them.  The success of the analytic mean
field theory is actually typical.  Although analytic approximations on
the TDGL itself have not been successful so far, transforming the
equation into an effective equation for the defects has proven to be the
appropriate starting point for semi-quantitatively accurate
predictions\cite{ojk}.

\subsection{The Kibble mechanism in the GUT transition}

It comes as a shock to most condensed matter physicists to learn that
cosmologists have a name for the mechanism in which topological defects
are formed during the quench through a phase transition.  (To my
knowledge, it has only occurred to one condensed matter physicist that
topological defects would {\it not\/} be created during a rapid
quench!)  During a putative phase transition in a grand unified theory
(GUT), string-like topological defects can be formed, which, many
expansion times later, have been proposed to act as seeds for galaxy
formation.  Although the initial string density does not turn out to be
important for existing models of structure formation, because the string
network is believed to attain a scaling solution at long times, some
attention has been given to the question of determining this quantity.

By the term \lq\lq initial string density" is meant the length of string
per unit volume \lq\lq just after the quench".  This only has a unique
meaning if the quench can be considered to be instantaneous: otherwise,
the order parameter configuration is determined by the history of the
system, and there is no unique value for this quantity.

Computer simulations and experiments on liquid crystals very clearly
show that following an instantaneous quench from above $T_c$, the order
parameter fluctuates wildly with position, and the order parameter
configuration is so disordered that it is meaningless to identify
topological defects.  After all, a topological defect is a defect
configuration in a smooth, ordered background, and when this does not
exist, individual defects cannot be distinguished.  This is equivalent
to saying that the cores of the defects overlap.  Thus, right after the
quench, the initial string density $\rho_i$ is not well-defined.  In
fact, it first becomes well-definable at a time $t_1$ when the average
separation $R(t_1)$ between defects is greater than the core size, given
by the order parameter correlation length $\xi (T_f)$ at the final
temperature $T_f$ reached by the quench.  Hence, the \lq\lq initial"
string density (meaning the string density at this time) can only be
given by
\be\label{initstr}
\rho_i = 1/R(t_1)^2 \sim \xi(T_f)^{-2}.
\ee
Note that the string correlation function at this and subsequent times,
which will depend on the quench history, as well as the static
correlations in the order parameter from the starting temperature above
$T_c$.

\subsection{Electroweak transition}

There is currently considerable interest in the dynamics of the electroweak
transition, due to the suggestion that
baryon asymmetry arises in the
propagating bubble walls accompanying this putative first order transition.
For a review, see the article by Turok in this volume.  The electroweak
transition is conceptually similar to the superconducting transition
when gauge field fluctuations are included\cite{halperin}, and we shall
couch our discussion in these terms.

The equilibrium statistical mechanics near the superconducting transition
is obtained from the partition function
\be\label{partsc}
Z = \int D\Psi\, D{\bf A} \, e^{-F\{\Psi, {\bf A}\}/k_BT}
\ee
where the coarse-grained free energy is usually taken to be
\be\label{glfe}
F\{\Psi, {\bf A}\}=\int d^d{\bf x}\,\left[ \frac{1}{2}|D\Psi|^2
+ {a}|\Psi|^2 + \frac{b}{2}|\Psi|^4
+ \frac{(\nabla\times {\bf A})^2}{8\pi} \right].
\ee
Here ${\bf A}$ is the electromagnetic vector potential, $D\Psi\equiv
\nabla - ie{\bf A}$ is the covariant derivative, and we have worked in
the gauge where the scalar potential is zero and $\nabla\cdot{\bf
A}=0$.  Halperin, Lubensky and Ma\cite{halperin} showed that the
functional integral over the gauge field leads to a term in the
resultant coarse-grained free energy density
\be\label{ftilde}
\tilde F\{\Psi\}
= -k_BT \,\log \int D{\bf A} \exp(-F\{\Psi, {\bf A}\}/k_BT)
\ee
which is proportional to $-|\Psi|^3$, and thus generates a first order
transition, which generically proceeds by nucleation and growth of
bubbles.  For strong supercooling, the nucleation barrier vanishes, and
the dynamics proceeds by spinodal decomposition.  In the cosmological
context of the electroweak transition, it is generally believed that the
transition occurs via nucleation and growth.  This occurs because the
Higgs mechanism causes the Higgs field to lose all degrees of freedom
apart from the modulus $|\Psi|$, and it is this quantity which appears in
the effective potential, and which varies across a domain wall.  There
are well-known problems associated with the computation of the effective
potential in this context, and the nature of the transition is currently
unclear.

The main remark that I would like to make here concerns the dynamics of
the phase transition.  In general, it is not correct to use the
derivative of $\tilde F\{\Psi\}$ as the driving force in a TDGL for
$\Psi$.  The reason is that the dynamics of $\Psi$ depends upon the
dynamics of the gauge field, and at least in the case of
superconductors, the relevant time scales for relaxation of both $\Psi$
and ${\bf A}$ are of the same order.  The correct procedure is to
coarse-grain both the microscopic order parameter and gauge field, and
then to solve the coupled dynamical equations for both $\Psi$ and ${\bf
A}$.  In the case of the superconductor transition, this has been done
in ref. \cite{lmg} with the result that growth does not proceed by
nucleation and growth of bubbles.  Elder and myself \cite{elder} have
repeated the calculation using the effective potential $\tilde F$, in
which the gauge field has been completely integrated out, and checked
that the results are quite different from those obtained using the full
dynamical equations for both the order parameter and gauge field.
A similar result was obtained by Ye and Brandenberger\cite{ye} in their
numerical simulations of the Abelian Higgs model, where no evidence for
domain wall formation was observed.  If this conclusion is supported by
further investigation, then there could be important ramifications for
proposed mechanisms of electroweak baryogenesis.

In conclusion, it seems that the time is ripe for a systematic study of
phase transition kinetics and the dynamics of fields, not only in the
context of condensed matter, but also in other areas where the
space-time behaviour of fields is of interest.  These include cosmology
and perhaps the physics of the quark-gluon plasma, soon to be probed by
RHIC.

\section{ACKNOWLEDGEMENTS}

The calculations of phase ordering referred to in the text were
performed in collaboration with M. Mondello, F. Liu, M. Zapotocky and
P.  Goldbart: it is a pleasure to acknowledge their contributions.  I
thank Ken Elder for many discussions on phase ordering in
superconductors, Yoshi Oono for his collaboration and many discussions,
Alan McKane for discussions on field theory, Andy Albrecht, Rob
Brandenberger, Anne Davis, Tom Kibble, Paul Shellard, Neil Turok,
Alexander Vilenkin and Erick Weinberg for numerous discussions on
cosmology, and the organisers of the workshop for successfully getting
condensed matter physicists and cosmologists to talk to each other.
Finally, I acknowledge the generous hospitality of the Isaac Newton
Institute, Cambridge.  This work was supported by National Science
Foundation grant NSF-DMR-93-14938.


\bb
\bibitem{mg}
M. Mondello and N.D. Goldenfeld, \journal Phys. Rev. A &42&5865(90).

\bibitem{mg3}
M. Mondello and N.D. Goldenfeld, \journal Phys. Rev. A &45&657(92).

\bibitem{mgcon}
M. Mondello and N.D. Goldenfeld, \journal Phys. Rev. E &47&2384(93); see
also, M. Siegert and M. Rao, \journal Phys. Rev. Lett. &70&1956(93).

\bibitem{lmg}
F. Liu, M. Mondello and N.D. Goldenfeld, \journal Phys. Rev. Lett.
&66&3071(91); H. Frahm, S. Ullah and A. Dorsey, \journal Phys. Rev.
Lett. &66&3067(91).

\bibitem{liu}
F. Liu, \journal Phys. Rev. E &48&2422(93).

\bibitem{zgg}
M. Zapotocky, P. Goldbart and N.D. Goldenfeld, {\sl Phys. Rev. E} (in
press).

\bibitem{toynum}
H. Toyoki, \journal Phys. Rev. A &42&911(90); \journal J. Phys. Soc.
Jpn. &60&1433(91); \journal ibid. &60&1153(91).

\bibitem{brayrev}
A.J. Bray, {\sl Adv. Phys.} (in press).

\bibitem{reviews}
For reviews on phase ordering see: J.D. Gunton, M. San Miguel and P.S.
Sahni in {\sl Phase Transitions and Critical Phenomena}, vol. 8, eds. C.
Domb and J.L. Lebowitz (Academic, New York, 1983), p. 267; K. Binder,
\journal Rep. Prog. Phys. &50&783(87); H. Furukawa, \journal Adv. Phys.
&34&703(85).

\bibitem{bray90}
A. J. Bray, \journal Phys. Rev. B & 41&6724(90).

\bibitem{tomita}
H. Tomita, \journal Prog. Theor. Phys. &75&482(86).

\bibitem{yeung}
C. Yeung, \journal Phys. Rev. Lett. &61&1135(88); H. Furukawa, \journal
Phys. Rev. B &40&2341(89); H. Furukawa, \journal J. Phys. Soc. Jpn.
&58&216(89).

\bibitem{porod}
G. Porod, in {\sl Small Angle X-ray Scattering}, edited by O. Glatter
and L. Kratky (Academic, New York, 1983).

\bibitem{coniglio}
A. Coniglio, P. Ruggiero and M. Zannetti, Scaling and Crossover in
the large-N model for growth kinetics, cond-mat/9405003.

\bibitem{shin}
A. Shinozaki and Y. Oono, \journal Phys. Rev. E &48&2622(93).

\bibitem{fetter}
See (\eg) the discussion by A.L. Fetter and J.D. Walecka, {\sl Quantum
Theory of Many-particle Systems} (McGraw-Hill, New York, 1971).

\bibitem{alford}
M. Alford and J. March-Russell, \journal Nucl. Phys. B &417&527(94).

\bibitem{wert}
For a review, see N.R. Werthamer, in {\sl Superconductivity}, edited
by R.D. Parks (Marcel Dekker, New York, 1969), vol. 1, p. 321.

\bibitem{enomoto}
Y. Enomoto, \journal Mod. Phys. Lett. B &5&1639(91).

\bibitem{toyoki}
H. Toyoki, in {\sl Dynamics of Ordering Processes in Condensed Matter},
ed. S. Komura, H. Furukawa (Plenum, New York, 1988).

\bibitem{pleiner}
H. Pleiner, \journal Phys. Rev. A &37&3986(88); P. Cladis \etal,
\journal Phys. Rev. Lett. &58&222(87); A. Pargellis \etal, \journal
Phys. Rev. A &46&7765(92); B. Yurke \etal, \journal Phys. Rev. E
&47&1525(93).

\bibitem{oonopuri}
Y. Oono and S. Puri, \journal Phys. Rev. Lett. &58&836(87); \journal
Phys. Rev. A &38&434(88); S. Puri and Y. Oono, \journal ibid.
&38&1542(88).

\bibitem{tom2}
H. Tomita, \journal Prog. Theor. Phys. &85&47(91).

\bibitem{rg}
See L.-Y. Chen, N.D. Goldenfeld and Y. Oono, \journal Phys. Rev. Lett.
&73&1311(94)
and references therein.

\bibitem{toymean}
H. Toyoki and K. Honda, \journal Prog. Theor. Phys. &78&237(87).

\bibitem{ojk}
T. Ohta, D. Jasnow and K. Kawasaki, \journal Phys. Rev. Lett.
&49&1223(82).

\bibitem{halperin}
B.I. Halperin, T.C. Lubensky and S.-K. Ma, \journal Phys. Rev. Lett.
&32&292(74).

\bibitem{elder}
K. Elder and N.D. Goldenfeld, unpublished.

\bibitem{ye} J. Ye and R. Brandenberger, \journal Nucl. Phys. B
&346&149(90).

\eb


\end{document}